# An Acoustic Method for Determining Ballistic Coefficients

Michael Courtney, PhD, Ballistics Testing Group, PO Box 3101, Cullowhee, NC, 28723
Michael_Courtney@alum.mit.edu

Amy Courtney, PhD, Department of Engineering, Western Carolina University, Cullowhee, NC, 28723  Amy_Courtney@post.harvard.edu

**Abstract:**
This paper presents a method for using a PC soundcard, microphone and a chronograph to determine bullet BC with an accuracy of 6%.  This is useful when a second chronograph is unavailable or when the projectile accuracy is insufficient to use a far chronograph.

**Introduction**
The ballistic coefficient (BC) describes how air resistance slows a projectile in flight [BAR97, BAR07, HOR91, NOS96, RES06, SPE94].  Earlier work has shown that manufacturer's BC numbers can be exaggerated by as much as 25% [COC07a].  An accurate BC is needed for shooting event reconstructions that use bullet penetration to determine distance or acoustic techniques to determine distance from target to shooter.  If an accurate BC is needed for a forensic investigation, it should be measured, preferably with the same firearm.

However, the method of determining BC by measuring near and far velocities is difficult to apply for shotgun pellets and projectiles without sufficient accuracy to reliably shoot through an optical chronograph at 50 or 100 yards.  We present a method for determining BC by measuring the near velocity via chronograph and the time of flight via acoustic measurement.  This technique is also useful in cases where only one chronograph is available.  (The far chronograph is often damaged by an errant bullet impact in the two-chronograph method.)

**Method**

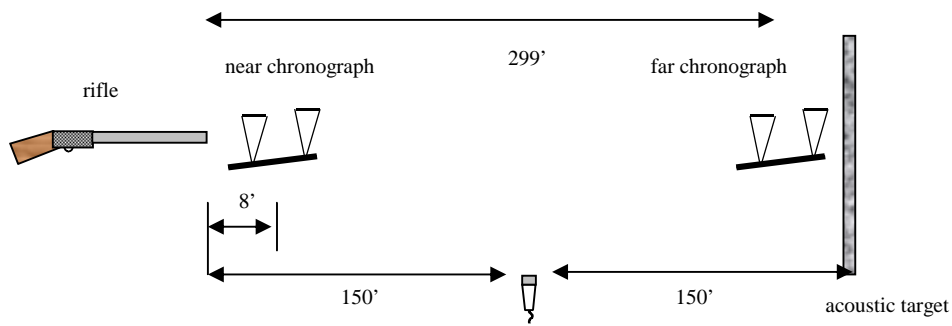

A resonating acoustic target is placed 300 feet from the muzzle.  An I beam is used here, but a much larger piece of metal may be used if necessary for accuracy reasons.  The key is mounting the target so that it makes a loud noise when the projectile hits.

A microphone is placed mid-way[1] between the muzzle and the acoustic target.  The sounds of the muzzle blast and target strike are easily detected on the digitized sound waveform.  Since the microphone is equidistant[2] from muzzle and target, sound takes the same time to reach the microphone from each source.  The time of flight is the same as the recorded time difference between these events.  The time of flight has an estimated uncertainty less than 50 microseconds.

A near chronograph is placed 8' from the muzzle to determine the near velocity.  The near velocity and the time of flight are used to determine the BC via the time of flight method.  A second chronograph is placed at 299 feet from the muzzle to independently determine the BC by direct measurement of the velocity loss.  This independent and more accurate determination of the BC is used as a check on the accuracy of the acoustic time of flight method.

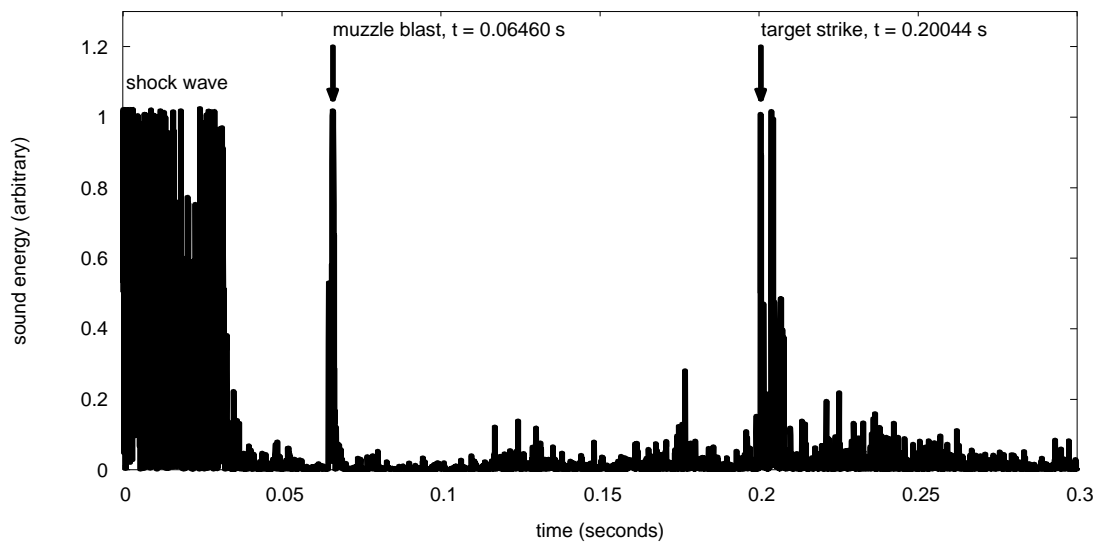

Figure 2: 223 Remington Sound Waveform

Figure 2 shows a sound waveform.  Sampling is triggered by the supersonic shock wave (t = 0) as the bullet passes the microphone.  The muzzle blast arrives next (t = 0.06460 s) followed by the sound of the bullet hitting the target (t = 0.20044 s).  The time of flight is the interval between the muzzle blast and target strike.

---

[1] The microphone needs to be equidistant from the muzzle and the target, it can be sufficiently to the side or underneath the projectile path so that risk of impact is small.
[2] Alternatively, one can place the microphone near the muzzle and subtract the time for the sound to travel back to the microphone from the target.  This introduces more uncertainty because it requires an accurate determination of the speed of sound under the ambient conditions.

The time of flight (0.13584 s), distance (300 ft), and the near velocity (2400 fps) can be used along with the air temperature (85 °F), atmospheric pressure (29.93 in Hg), relative humidity (41%), and altitude (1100 ft) to determine the bullet BC [JBM07].

**Results and Discussion**
Simultaneous measurements with both the acoustic and the near/far velocity methods were conducted on five consecutive shots from a .223 rifle.  Acoustic time of flight BC determinations are compared with the more accurate technique of near/far velocity measurements below:

| Shot | BC (velocity) | BC (time) | difference (%) |
|---|---|---|---|
| 1 | 0.215 | 0.207 | -3.72% |
| 2 | 0.218 | 0.239 | 9.63% |
| 3 | 0.216 | 0.213 | -1.39% |
| 4 | 0.217 | 0.218 | 0.46% |
| 5 | 0.224 | 0.244 | 8.93% |
| Mean | 0.218(3) | 0.224 (14) | 2.75% |

Numbers in parenthesis represent the estimated uncertainty in the last significant digit(s) of the reported mean BC.

The average BC value from the acoustic method is within 3% of the value determined from direct measurement of velocity loss, but the estimated uncertainty of the acoustic measurement is 6.5%.  This level of uncertainty will be sufficient for many purposes, but can be reduced simply by increasing the number of shots.   The Hornady reloading manual [HOR91] lists the BC of this bullet as 0.235, but many manufacturer specifications for BC are exaggerated.

Most of the uncertainty in the acoustic technique results from the time response of the microphone and sample rate of the digitization.  Much smaller uncertainty would result from attaching piezoelectric transducers both to the muzzle and to the acoustic target and sampling both waveforms at 1 MHz.  This would give an uncertainty in the time of flight on the order of 2 microseconds and much more accurate BC determinations for a relatively small number of shots.  However, this requires a significant investment in expensive electronic equipment rather than a simple $200 chronograph and a notebook PC with a soundcard.

**References:**
[BAR97] **Barnes Reloading Manual Number 2-Rifle Data**, 1997, Barnes Bullets, Inc., American Fork, UT.
[BAR05] www.barnesbullets.com accessed 12/20/1995.
[BAR07] www.barnesbullets.com accessed 2/2/2007.

[BER07] www.bergerbullets.com accessed 2/2/2007.
[COC07a]   Courtney M, Courtney A, The Truth About Ballistic Coefficients, (2007).  Search title at arxiv.org.
[HOR91] **Hornady Handbook of Cartridge Reloading**, Fourth Edition, 1991, Hornady Manufacturing Company, Grand Island, NE.
[HOR07] www.hornady.com accessed 2/2/2007.
[JBM07]    http://www.eskimo.com/~jbm/calculations/calculations.html   accessed 1/30/2007.
[NOS96] **Nosler Reloading Guide Number Four**, 1996, Nosler, Inc., Bend OR.
[NOS07] www.nosler.com accessed 2/2/2007.
[RES06]  http://www.shootingsoftware.com/coefficients.htm accessed 6/25/2006.
[SPE94]  **Speer Reloading Manual Number 12**, 1994, Blount, Inc., Lewiston, ID.
[WIN07] www.winchester.com accessed 2/2/2007.